# Reversible Engineering of Topological Insulator Surface State Conductivity through Optical Excitation

Faji Xie[1,*], Zhen Lian[2,*], Shuai Zhang[1,*], Tianmeng Wang[2], Shengnan Miao[2], Zhiyong Song[3], Zhe Ying[1], Xing-Chen Pan[1], Mingsheng Long[3], Minhao Zhang[1], Fucong Fei[1], Weida Hu[3], Geliang Yu[1], Fengqi Song[1,†], Ting-Ting Kang[3,‡], and Su-Fei Shi[2,4, §]

1. National Laboratory of Solid State Microstructures, School of Physics and Collaborative Innovation Center of Advanced Microstructures, Nanjing University, Nanjing 210093, China
2. Department of Chemical and Biological Engineering, Rensselaer Polytechnic Institute, Troy, New York 12810, USA
3. State Key Laboratory of Infrared Physics, Shanghai Institute of Technical Physics, Chinese Academy of Sciences, Shanghai 200083, China
4. Department of Electrical, Computer & Systems Engineering, Rensselaer Polytechnic Institute, Troy, New York 12180, USA

**Despite the broadband response, limited optical absorption at a particular wavelength hinders the development of optoelectronics based on Dirac fermions[1,2]. Heterostructures of graphene and various semiconductors have been explored for this purpose[3], while non-ideal interfaces often limit the performance. The topological insulator is a natural hybrid system[4-6], with the surface states hosting high-mobility Dirac fermions and the small-bandgap semiconducting bulk state strongly absorbing light. In this work, we show a large photocurrent response from a field effect transistor device based on intrinsic topological insulator Sn-$Bi_{1.1}Sb_{0.9}Te_2S$. The photocurrent response is non-volatile and sensitively depends on the initial Fermi energy of the surface state, and it can be erased by controlling the gate voltage. Our observations can be explained with a remote photo-doping mechanism, in which the light excites the defects in the bulk and frees the localized carriers to the surface state. This photodoping modulates the surface state conductivity without compromising the mobility, and it also significantly modify the quantum Hall effect of the surface state. Our work thus illustrates a route to reversibly manipulate the surface states through optical excitation, shedding light into utilizing topological surface states for quantum optoelectronics.**

---

* The first three authors contributed equally.
† Corresponding author. Emails: songfengqi@nju.edu.cn.
‡ Corresponding author. Emails: kang@mail.sitp.ac.cn.
§ Corresponding author. Emails: shis2@rpi.edu.

Dirac fermions[7,8] with high mobility have been widely explored for electronic devices[1,2], and it was also proposed for optoelectronics for its broadband absorption[9,10]. However, monolayer graphene only absorbs about 2.3% light in the visible to midinfrared regime[11,12], limiting the possible applications. Various combinations of semiconductor and graphene heterostructures[13-17], therefore, have been explored to overcome this limitation, with the semiconductor absorbing light and the graphene hosting high mobility conducting channels. In these heterostructures, light excited carriers need to be transferred efficiently to the graphene layer, which are often constrained by non-ideal interfaces. Topological insulator[4-6] (TI), with its symmetry protected surface states and semiconducting bulk state, provides a natural hybrid structure to bypass this dilemma. The surface states of the TI possess spin-polarized linear dispersion that hosts massless Dirac fermions, similar to graphene but with spin degeneracy lifted. The bulk state of the topological insulator is a small bandgap semiconductor, which strongly absorbs light in the visible to midinfrared regime. However, the direct optical manipulation of the quantum transport behaviors of the topological surface state (TSS) is still lacking, because of the conductivity of TSS are often shadowed by the bulk conductivity due to unintentional doping.

In this work, we investigate the photo-response of a field effector transistor (FET) based on intrinsic TI: Sn-$Bi_{1.1}Sb_{0.9}Te_2S$ (Sn-BSTS). We found a large and non-volatile photocurrent response from the TI FET device, which arises from the effective doping of the TSS. The photocurrent response sensitively depends on the initial Fermi energy of the TSS, and the memory effect can be erased by a large gate voltage or raising the temperature. We further propose a remote doping mechanism, in which the light excites the defect states in the bulk and the trapped carriers are freed to the TSS. The resulting photodoping leads to the greatly modified the conductivity of TSS observed in the experiments. Our interpretation is further confirmed by the quantum transport measurements under the continuous illumination of light. The photodoping modifies the quantum Hall effect (QHE) in the TI FET by broadening and shifting the position of the quantum Hall conductance plateau. Our study shows the possibility of using light excitation as a new knob to manipulate the TSS of TI in the quantum Hall regime.

**Results**
**Photodoping of TSS of an intrinsic topological insulator**

The TI FET device is schematically shown in Fig. 1a. The TI flake is typically ~100 nm thick, placed on 300 nm thermal $SiO_2$ and gated by the silicon back gate (see Methods section). In dark, the electrical transport measurement of the device at 2 K clearly shows a bipolar behavior (Fig. 1b, black curve), a signature of the TSS dominated transport. The optimized TI, Sn-BSTS, has been reported to have a mobility up to 10,277 $cm^2$/Vs when the Fermi energy is tuned close to Dirac point[18]. The minimum conductivity of the device shown in Fig. 1 is ~ 4.5 $G_0$, consistent with previous report[19,20]. The charge neutral point (CNP) is close to zero gate voltage (~ -5.5 V), suggesting that the Fermi energy is not only in the bandgap but also very close to the Dirac point of the TSS, a result of minimum initial doping of the surface state. We then illuminate the device with a LED light source, with excitation power of 11 µW and center wavelength of 945 nm. The resulting transfer curves with the light on maintain the bipolar behavior, suggesting that the transport is still dominated by the TSS (later confirmed by quantum Hall measurements). However, there are two major changes: (1) The CNP shifts to ~ -59.5 V; (2) the conductance slope becomes flatter (Fig. 1b). The shift of CNP can be explained with the effective electron doping of the TSS upon optical excitation. The flatter conductance, however, is not due to the decreased mobility but decreased effective capacitance (Supplementary Information section 4), as we discuss later.

These changes lead to a large photocurrent response, ~ -0.256 nA for both CNP (-5.5V) and gate voltage of 60 V, with a bias as small as ~3 mV. The photocurrent sign change is a natural consequence of photo-induced conductivity change of the TSS conductivity (Fig. 1b). As shown in Fig. 1b, the photocurrent is positive in region II (-25.0 V to 29.0 V) but negative in region I and III. Time-dependent photoconductivity measurement is also performed at fixed $V_g$ to confirm the gate tunable transition between increased and deceased photoconductivity (Supplementary Information section 3).

The gate voltage in Fig. 1b are swept from 60 to -80 V, as indicated by the arrow direction in Fig. 1b. We further measure the transfer curves by sweeping the gate voltage on both directions. As shown in Fig. 1c, in dark, the TI device shows a large hysteresis (black curve). Previously, the origins of the hysteresis in graphene and transition metal dichalcogenides FETs have been systematically studied and are attributed to trap states or mobile ions inside gate dielectric layers or water molecules trapped in graphene/oxide interfaces[21-24], which tend to be suppressed either in high vacuum or at low temperature[22-24], opposite to our observations of hysteresis at temperature of 2 K. Our controlled device of h-BN gated TI devices show similar behavior of hysteresis upon light excitation (Supplementary Information section 7), which indicates that the hysteresis is an intrinsic properties of Sn-BSTS[25-27]. Interestingly, the hysteresis decreases as the excitation power increases and eventually disappears for the LED power of 11 µW (red curves in Fig. 1c).

**Photodoping mechanism**

The transfer curves of our device measured at different excitation powers are shown in Fig. 2a. As the excitation power increases, the effective doping of the TSS increases, as evidenced by the shift of CNP. The amount of photodoping can be estimated from the shift of CNP by the expression $\Delta n = c_{\text{eff}} \Delta V_{\text{CNP}}/e$, where $c_{\text{eff}}$ is effective gate capacitance per unit area, $\Delta V_{\text{CNP}}$ is the shift of CNP and $e$ is the electron charge constant (Supplementary Information section 5). It can be found that the photodoping of TSS increases as a function of the LED power at low excitation power, while it saturates around 5.6 µW. At the excitation power 11 µW, the effective electron doping is $\Delta n_e = 3.3 \times 10^{11}$ cm$^{-2}$, which roughly correspond to a Fermi energy shift of 26.8 meV.

Interestingly, the minimum conductivity remains roughly the same (~4.5 $G_0$) for different optical excitation powers. This insensitive dependence of the minimum conductivity on increased charge impurity scattering center on the surface was reported in graphene previously[19,28], but our observation is the first report on TSS of TI. The light excitation, therefore, could be controlled to investigate the minimum conductivity of Dirac fermions of TI.

The photodoping effect is non-volatile and persists after turning off the light (Supplementary Information section 3). Further, the photodoping effect sensitively depend on the initial Fermi level of the TSS. To check this behavior in detail, we take a different approach of illuminating the second device (Fig. 2b). Here, we turn on the optical excitation of 11 μW at one particular gate voltage. We then turn off the light and sweep the gate voltage to -80 V. The data for different initial gate voltages are shown in Fig. 2b. Compared with the transfer curve taken in the dark (dashed black line in Fig. 2b), the flashed illumination at the gate voltage of 60 V shows a transfer curve of a p-doped TSS. With the decrease of the initial gate voltage that we flash the light, the CNP voltage keeps decreasing, and the transfer curve eventually turns to one corresponding to the n-doped TSS.

All our observations can be explained with a remote photo-doping mechanism. Shown in Fig. 2c-e, there are a distribution defect states in the bulk that are close to the Direct point of TSS and act as donor-like and acceptor-like states (only show two discrete levels for simplicity in the schematics) to the TSS that can be activated through light. Upon illumination, the trapped electron or hole can be activated to the high energy conduction band or valence band, and then relax to the low energy state, either the TSS or the original defect states. When the initial gate voltage is at 60 V for the flashed light excitation, TSS is n-doped. Optically excited electron would prefer to relax to the donor-like defects state and then be stuck due to the large binding energy of the localized state. However, the optically excited holes from the acceptor-like defects can transfer to TSS, leaving behind negatively charged defects that effectively gate the TSS to be more p-doped compared with the original TSS not exposed to light. Once the holes transferred to the TSS, they quickly relax energy and cannot go back to the defect states, leading to the non-volatile change after the light is turned off. As the initial gate voltage for the flashed light excitation decreases, TSS is less n doped. The Fermi energy is lower than some of the donor-like defect states, and optically excited electron start to transfer to TSS. As a result, the TSS start to be more n-doped. For the initial gate voltage close to CNP (Fig. 2d), almost all optically excited electrons and holes can transfer to TSS. There are more donor-like defect states the acceptor-like defect states, and as a results, there are more positively charged defects left behind after the carrier transfer, which act as the effective gating that n-dope the TSS. Finally, as the initial gate voltage was tuned to -30 V, light excited holes transfer to TSS start to get blocked, while all the optically excited electron can be transferred to the TSS, leading to even more n-doped TSS after the flashed optical excitaiton.

It is worth noting that there are two ways to erase the non-volatile doping effect (Supplementary Information section 2). One is to sweep $V_g$ below -80 V or larger than 60 V. This suggests that at corresponding Fermi energies, the carrier have enough energy to go back to defect states and then trapped there. Considering the effective capacitance, this suggests that the defect states are distributed around the Dirac point by ±(30-40meV) (Supplementary Information section 8). The other way is to raise the temperature to be over 100 K and then cool the device down to 2 K. This suggests that the thermal excitation of ~ 10 meV is enough to facilitate the carrier go back to defect states, which is more effective than tuning the gate voltage. The reason is that, according to the Fermi distribution, there is a small population of carrier with the kinetic energy larger than 10 meV. Considering the fast tunneling time and finite electrical transport measurement time (10s'), even a finite tunneling probability would drive all carriers to defect states during our measurement. The origin of these defects is beyond the scope of this work but certainly warrants future exploration considering the intense interest in intrinsic TIs.

Notably, with this model, we can also explain the decrease of the gate efficiency upon light illumination which flattens the transfer curve (Fig. 1b), which is also later confirmed by QHE measurement (Supplementary Information section 4). Since the optical excitation creates empty defect states, the carriers induced by the gate voltage have to fill the defect states before filling the TSS.

**Optically tunable QHE of TSS**

Magneto-transport measurements have been employed to unambiguously demonstrate QHE of the TSS[18,20,29-31] and associated nontrivial Berry phase[32]. Here we measure the hall conductivity of the TI device with and without the continuous optical excitation. The gate voltage dependence of longitudinal conductance $G_{xx}$, Hall conductance $G_{xy}$, as well as two-terminal conductance $G_{2\text{-term}}$ are measured at $B = 14$ T and $T = 2$ K, all exhibiting the characteristics of TSS-dominated transport (Fig. 3a-c). In dark, $G_{xy}$ shows two well-developed plateaus, which correspond to the total Landau filling factor $\nu = 0$ and -1, respectively (Fig. 3b). The transition between the two plateaus, accompanied by one peak in $G_{xx}$, is located very close to the CNP (Fig. 3c), indicating the filling of zeroth Landau level of the bottom surface. Such observation is in good agreement with previous studies[18,20,31] and can be explained by half-integer QH conductivity from both the top and bottom surface states: while the filling factor of the top surface $\nu_{\text{top}}$ remains as -1/2, $\nu_{\text{bottom}}$ can be tuned from -1/2 to 1/2 by the back-gate voltage.

We further examine the influence of the photodoping effect on the QHE. Upon the continuous LED light illumination with the power 11 μW, it is evident from Fig. 3b that the QHE plateaus of $-e^2/h$ of $G_{xy}$ shifts to the negative gate voltage and also broadens in width, and the corresponding $G_{xx}$ peak exhibit a shift in position as well. As shown in Fig. 3c, the zeroth Landau level (LL) has shifted from -6.6 V to -65.8 V, which is the same as the shift of CNP for the device 1 as shown in Fig. 1b. These results show that, upon light illumination, the transport of the TI device shows QHE of an electron doped TSS, as schematically shown in Fig. 3e.

The analysis of QHE is given by renormalization group flow diagrams (RGFDs)[33,34] as shown in Fig. 3d and Methods section, which indicates complete Landau quantization of the bottom TSS. For RGFDs under light illumination, the sets of converging points can be observed with $G_{xy}$ pointing to 0 and $-e^2/h$. It is clearly that all the RGFDs under different excitation power show continuous evolution of Landau quantization and reach an integer. It indicates that TSS is still a well-protected 2D electron system under light illumination. We further confirmed that, through analysis of weak anti-localization, transport dissipation from bulk can be ignored (Supplementary Information section 6).

In summary, we report the observation of photo-doping of the TSS of a FET device based on Sn-BSTS, and the effects can be sensitively tuned by the excitation power and gate voltage. The gate voltage controlled non-volatile photodoping of TSS promise applications in novel memory devices. Since the photodoping arise from the activation of the defect states in bulk and their coupling to the TSS, our study also inspires future work on defect engineering in TI bulk state for optoelectronics based on TSS. Further, the optical excitation of these defects can be utilized as a new knob to explore quantum transport of TSS such as the origin of minimum conductivity and optically controllable QHE.

**Methods**
**Crystal growth and device fabrication.** High-quality Sn-BSTS crystals were grown by melting elements of Bi, Sb, Te, S in a molar ratio of 1.1:0.9:2:1 and a small quantity of Sn at 750 ℃ for 24h in evacuated vacuum quartz tubes, followed by cooling to room temperature naturally. The crystal with a low bulk carrier density can be cleaved easily. With mechanical exfoliation, the Sn-BSTS flakes were transferred onto $SiO_2$/Si substrates, which was then fabricated as gated Hall-bar devices by photolithography, electron-beam evaporation and standard lift-off techniques.

**Transport measurements and analysis.** The transport measurements were carried out by a physical property measurement system at low temperatures down to 2 K with a magnetic field up to 14 T. We measured the resistances by standard lock-in amplifiers (SR830) with a low-frequency (13.33Hz) excitation current of 1uA (Keithley 6221). A direct current source (Keithley 2400) was used to power the LED. The conductivities are calculated by $G_{xx} = R_{sh}/(R_{sh}^2 + R_{xy}^2)$ and $G_{xy} = R_{xy}/(R_{sh}^2 + R_{xy}^2)$, $R_{sh} = R_{xx}(W/L)$. $G_{xx}$ and $G_{xy}$ are dependent on the parameters such as temperature, magnetic field, back-gate voltage and light illumination. The RGFDs allow extraction of two sets of converging points in the ($G_{xy}$, $G_{xx}$) space. Here the temperature and magnetic field are fixed, and the RGFDs are plotted with different light excitation power for different back-gate voltages. Each curve in the RGFDs is under a fixed light illumination power (dark as the excitation power of zero), which corresponding to different doping and can be viewed as different Fermi levels.

**Acknowledgements**

We gratefully acknowledge the financial support of the National Key R&D Program of China (2017YFA0303203), the National Natural Science Foundation of China (91622115, 11522432, 11574217, U1732273, U1732159, 61822403, 11874203, 11904165 and 11904166), the Natural Science Foundation of Jiangsu Province (BK20160659 and BK20190286), the Fundamental Research Funds for the Central Universities, and the opening Project of the Wuhan National High Magnetic Field Center. We also acknowledge the assistance of the Nanofabrication and Characterization Center at the Physics College of Nanjing University. Z.L., T.W., S.M. and S.-F. Shi acknowledges the support from AFOSR through Grant FA9550-18-1-0312, ACS PRF through Grant 59957-DNI10, NYSTAR through Focus Center-NY–RPI Contract C150117, and NSF through Career Grant DMR-1945420.

**Author contributions**

FS conceived the work. FX fabricated the devices and performed the transport measurement as assisted by SZ. X.-C.P. and F. F. prepared the bulk crystal. ZS, ML, WH, ZL, S.-F.S. and T.-T.K. provided the conditions of light experiment. FX, ZL, TW, SM and S.-F.S. proposed the photo doping mechanism model. FX, ZL, S.-F.S and FS co-wrote the paper. SZ, MZ, and GLY participated in the discussions. All authors commented on the manuscript.

**Additional information**

Correspondence and requests for materials should be addressed to FS.

**Competing financial interests**

The authors declare no competing financial interests.

**Figures**

**Figure 1. Light induced doping of surface states of an intrinsic topological insulator.** (a) Schematic of topological insulator Sn-BiSbTeS(1.1:0.9:2:1) based phototransistor. Purple lines indicate the topological surface states (b) Comparison of Transfer curves for dark and 11 μW light illumination at 2 K. (c) Transfer curves with hysteresis under different optical excitation power. The arrows indicate the scanning direction.

**Figure 2. Controllable photodoping of the topological surface state.** (a) Transfer curves for different optical excitation powers at 2 K. (b) Transfer curves displaying non-volatile photodoping effect for different initial gate voltages under which the flashed light illumination occur. (c)-(e) Schematics showing the photodoping mechanism for initially n-doped (c), intrinsic (d), and p-doped TSS.

**Figure 3. QHE of TSS upon light illumination.** (a) Back gate voltage dependent two-terminal conductance, $G_{2\text{-term}}$, at 14T for different optical excitation powers. The dashed lines indicate QHE states with index $v$. The inset shows the measurement configuration. The arrows show direction of sweeping back-gate voltage. (b) and (c) are $G_{xy}$ (b) and $G_{xx}$ (c) as a function of the back gate voltage at 14T for different light illumination power, respectively. (d) The renormalization-group flow diagram (RGFD) in ($G_{xy}$, $G_{xx}$) space. Each curve in the RGFD correspond to one fixed light illumination power (Dark, 0.01 μW, 0.1 μW, 0.5 μW, 1.1 μW, 5.6 μW, and 11 μW). (e) Schematics showing the Fermi energy of the TSS with respect to the LLs for both dark and upon illumination, with the back gate voltage set at -30 V.

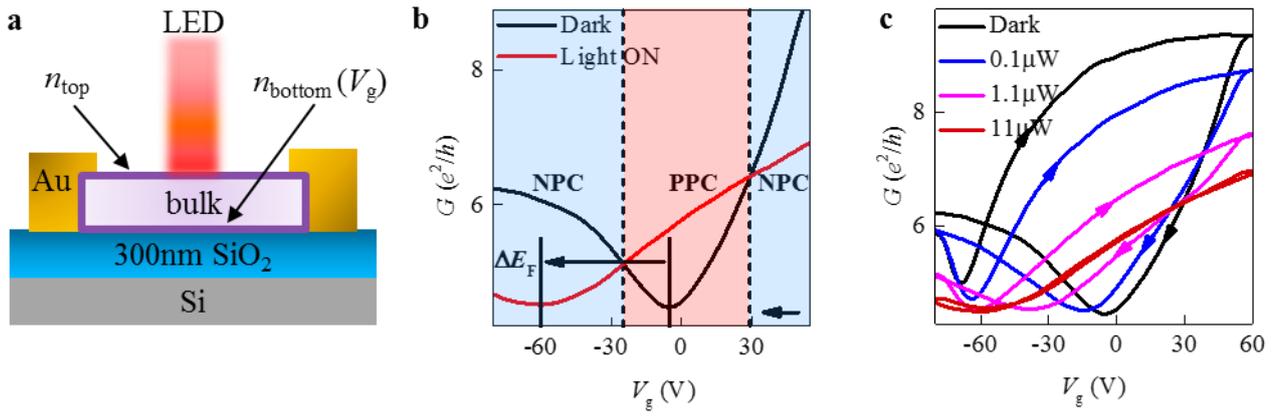

**Figure 1**

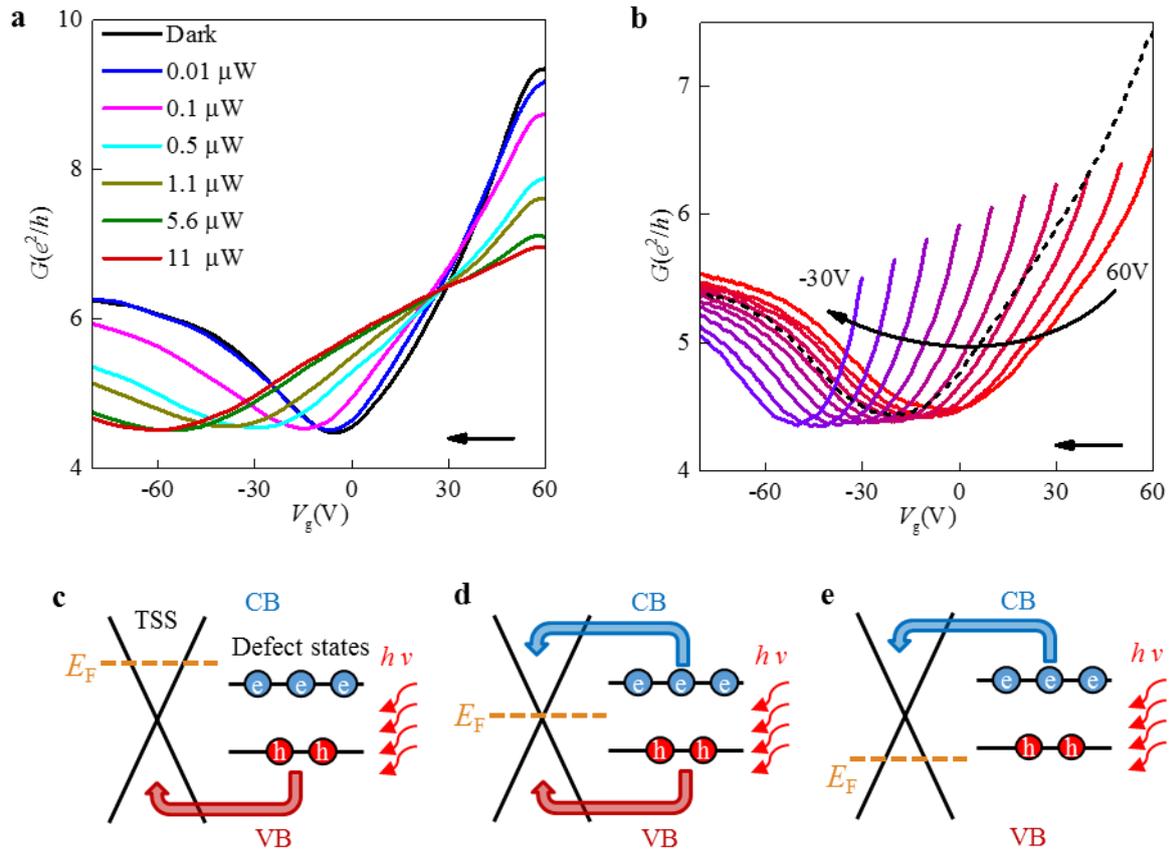

**Figure 2**

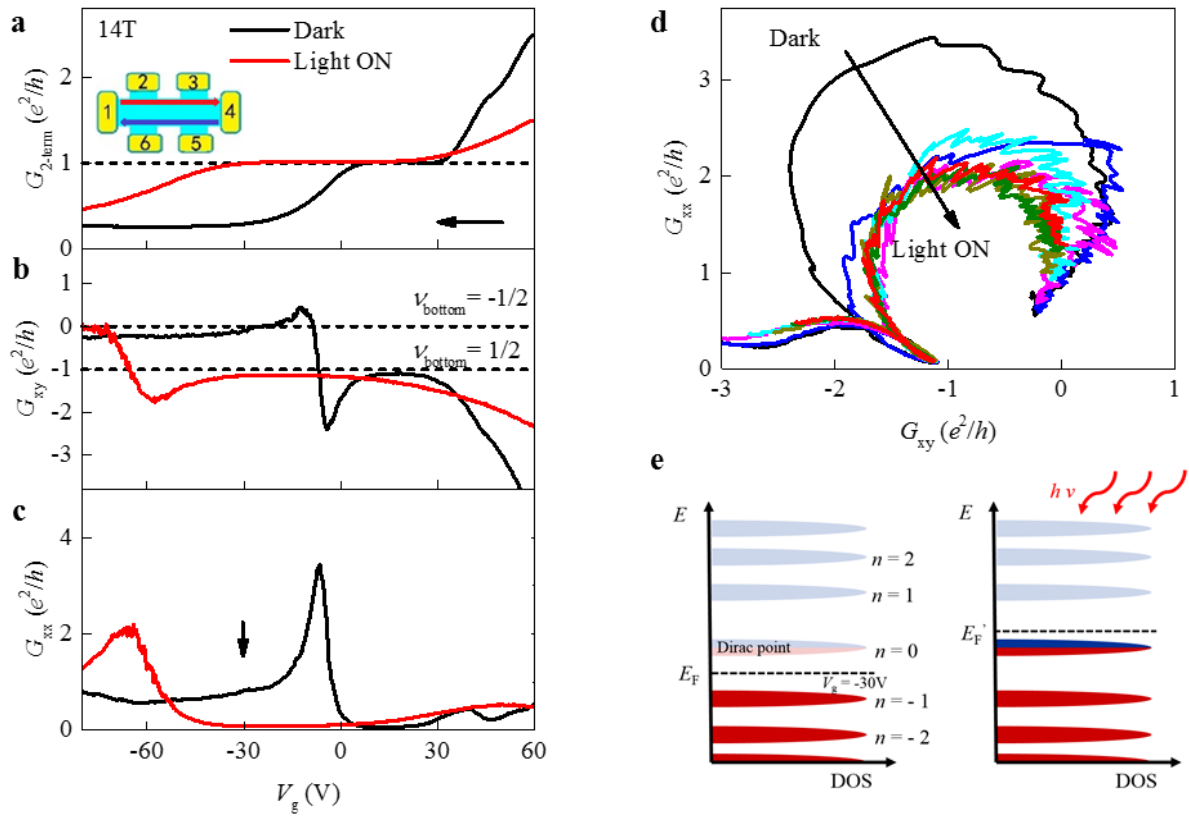

**Figure 3**

**Supplementary Information**
**Reversible Engineering of Topological Insulator Surface State Conductivity through Optical Excitation**

Faji Xie[1], Zhen Lian[2], Shuai Zhang[1], Tianmeng Wang[2], Shengnan Miao[2], Zhiyong Song[3], Zhe Ying[1], Xing-Chen Pan[1], Mingsheng Long[3], Minhao Zhang[1], Fucong Fei[1], Weida Hu[3], Geliang Yu[1], Fengqi Song[1], Ting-Ting Kang[3], and Su-Fei Shi[2,4]

1. National Laboratory of Solid-State Microstructures, School of Physics and Collaborative Innovation Center of Advanced Microstructures, Nanjing University, Nanjing 210093, China
2. Department of Chemical and Biological Engineering, Rensselaer Polytechnic Institute, Troy, New York 12810, USA
3. State Key Laboratory of Infrared Physics, Shanghai Institute of Technical Physics, Chinese Academy of Sciences, Shanghai 200083, China
4. Department of Electrical, Computer & Systems Engineering, Rensselaer Polytechnic Institute, Troy, NY 12180, USA

**Contents**
**1. Device picture**
**2. Temperature and gate voltage tuned charge memory**
**3. Time dependent conductivity curves with different chemical potential**
**4. Effective gate capacitance**
**5. Effective photo-doping of TSS**
**6. Weak anti-localization analysis**
**7. Boron nitride gated devices**
**8. Calculation of the Fermi energy of the TSS**

## 1. Device picture

As shown in Fig. S1b, using mechanical exfoliation, Sn-BSTS nanoflakes were transferred to $SiO_2$/Si substrates before being fabricated into gated Hall-bar devices. An atomic force microscopy image of the device is shown in Fig. S1a. The thickness of the device is ~ 100 nm, extracted from Fig.S1a.

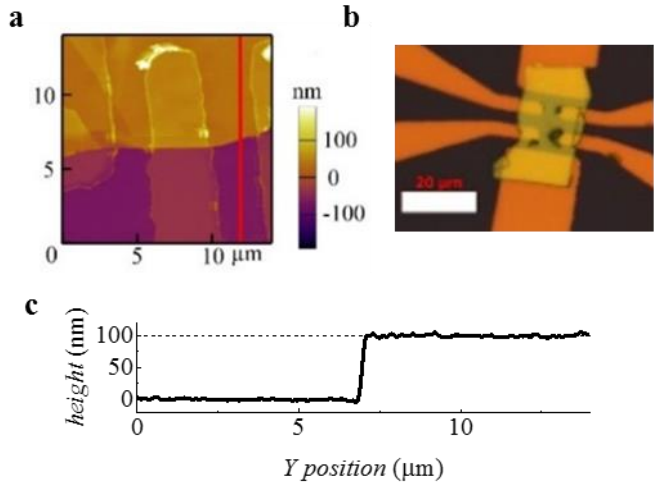

**Figure S1.** (a) Atomic force microscopy image of a typical Hall-bar device. (b) The optical microscope image of the device. scale bar: 20 μm. (c) Sample height profile along the red line marked in (a).

## 2. Temperature dependence and erasing the nonvolatile photodoping effect.

For n-doped TSS, the photo-induced conductance change is obvious for low temperature up to 50 K, while it disappears rapidly above 60 K, as shown in Fig. S2b.

As shown in Fig. S2c, in the temperature range of 200-300 K, the dark conductance $G(T)$ curve (red line) shows a bulk insulating phase behavior. When the temperature is below 200 K, the transport curves show metallic behaviors, implying that surface states starts to dominate. When the temperature is lower than 25 K, only topological Dirac fermion from TSS contribute to the conductance. After a flash of 11 μW excitation power, we wait for the conductance in dark (second dark) to be stable for 2 hours and then test the temperature dependent conductance curve (black line). From the difference between red and dark lines in Fig. S2c, the nonvolatile photodoping effect (memory effect) completely disappears above 100K.

Another method to erase the nonvolatile doping effect (Fig. S2d) at 2 K is by a circling the gate voltage from -80 V to 60 V.

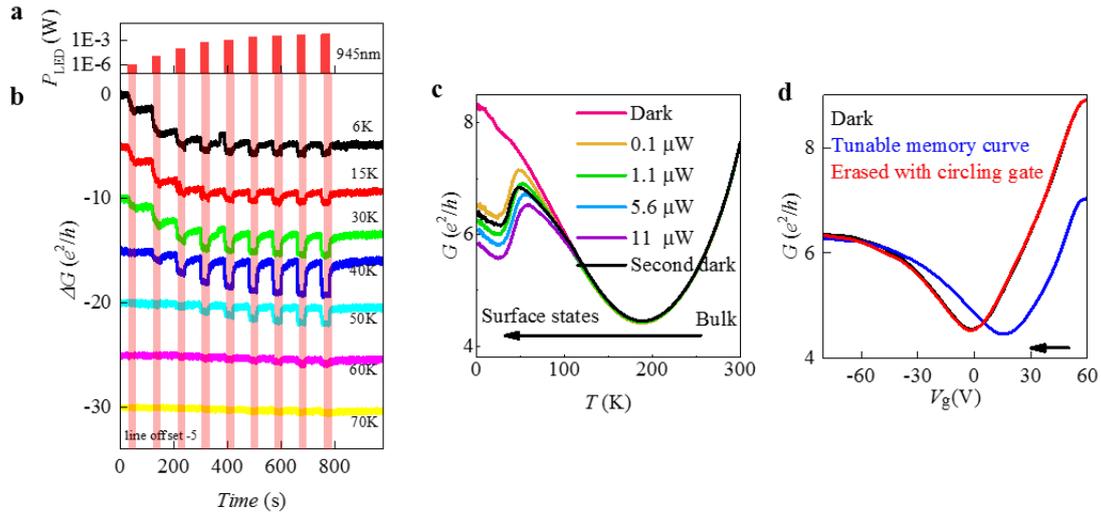

**Figure S2.** (a) Time-dependent LED excitation power as light is turned on and off. The LED excitation power is set to 0.01 μW, 0.1 μW, 1.1 μW, 5.6 μW, 11 μW, 21 μW, 31 μW, 42 μW and 53 μW sequentially for the light centered at 945 nm. (b) The change of conductance following the illumination sequence for different temperatures, with the offset of -5$e^2$/h. (c) The typical temperature-dependent conductance under different LED light illumination powers. The second dark curve is measured after turning off the LED for 2 hours. (d) Transfer curves under different photoinduced doping operation. Black: the original transfer curve indicates typical bipolar transport of TSS; bule: a selected tunable memory curve with CNP shifted; red: the transfer curve is measured after erasing nonvolatile doping memory by sweeping gate to -80 V and back to 60 V.

## 3. Time dependent photodoping effect for different doping of TSS

Time-dependent photoconductivity measurement is also performed in Fig. S2a and S2c at different fixed back gate voltages $V_g$. When the initial Fermi energy of TSS is close to Dirac point at dark condition (region II), the photo-induced conductance change is positive. When the chemical potential of TSS is in region I and III, the photo-induced conductance change is negative (Fig.S3b).

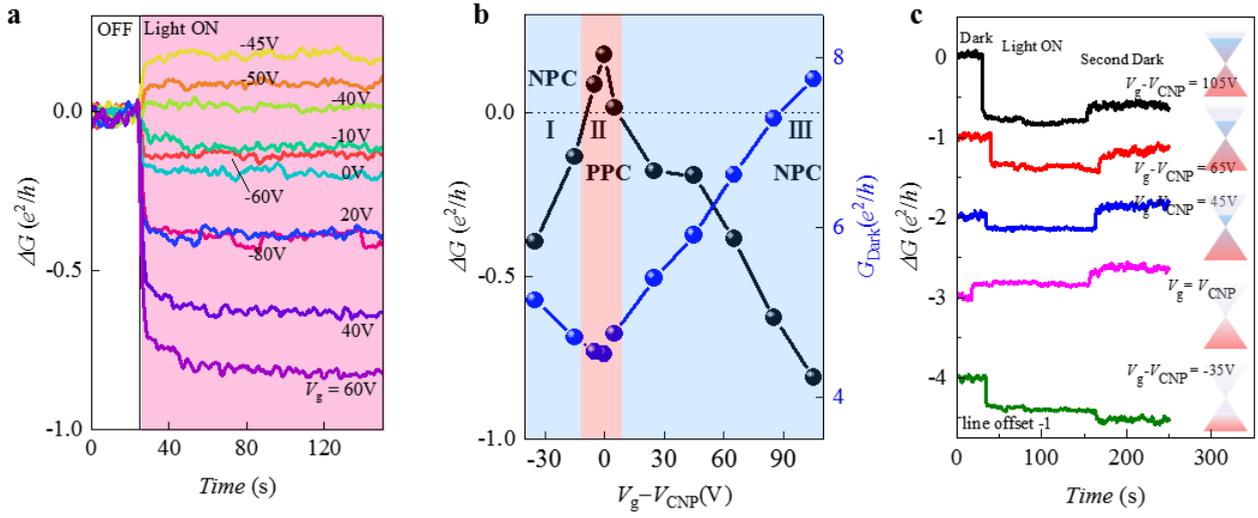

**Figure S3.** (a) Time dependence of conductivity change under different back-gate voltage. (b) Effective gate voltage dependence of photo-induced conductance change (dark dots) and dark conductance (blue dots), $G_{dark}$, extracted from (a). (c) Photo-induced conductance change as a function of time for different back gate voltages. The first and second abrupt change of conductance corresponds to the time to turn on the light and turn off the light, respectively.

## 4. Effective gate capacitance

We also measured the effective gate capacitances in dark and under light illumination with Hall measurements at low magnetic field. As shown in Fig. S5a and S5c, the effective gate capacitance decreases from 20% to 8.5% with 11 μW light illumination. The 100% efficiency correspond to the geometry gate capacitance of the 300 nm $SiO_2$. The decreased gate efficiency under light illumination indicates that additional carriers will be needed to fill the defect states.

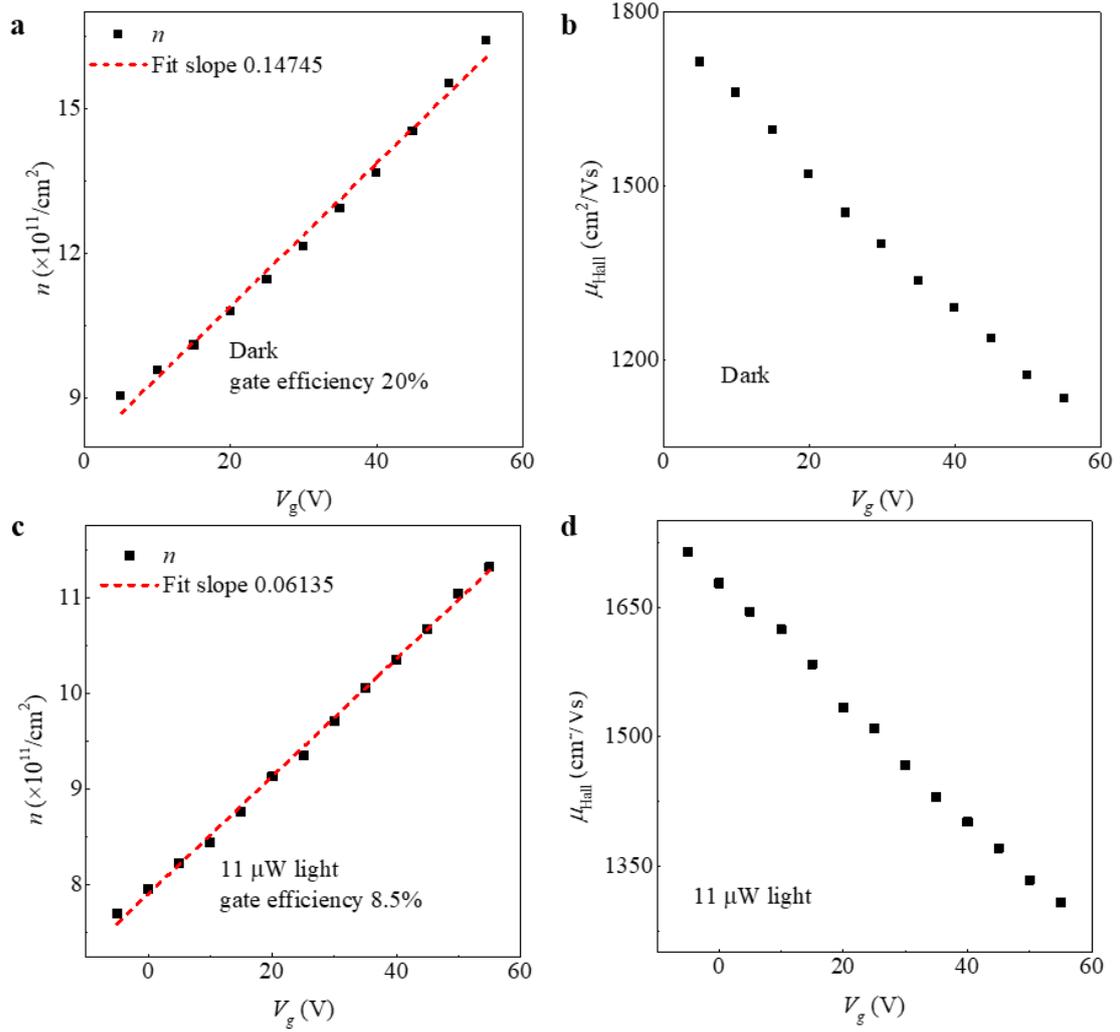

**Figure S4.** (a) and (c) are the changes of carrier density under different fixed back gate voltage in dark and 11 μW light. Red dashed lines are the linear fittings. (b) and (d) are the Hall mobility corresponding to the carrier density in a and c, respectively.

## 5. Effective photodoping of TSS

As the excitation power increases, the effective doping of the TSS increases, as evidenced by the shift of CNP. The amount of photodoping can be calculated from the shift of CNP with $\Delta n = c_{\text{eff}} \Delta V_{\text{CNP}}/e$, where $\Delta V_{\text{CNP}}$ is the shift of CNP, $e$ is the electron charge constant, and $c_{\text{eff}}$ is effective gate capacitance under different light illumination conditions. The effective gate capacitance for the condition the dark and illumination of power of 11 µW can be extracted from Fig. S4 seperately, and we extrapolate to obtain the gate capacitance for the illumination power less than 11 µW. The results are shown in Fig. S5. It can be found that $\Delta n$ increases with $P_{\text{LED}}$ at the low excitation power, while it star to saturates at the illumination power of 5.6 µW, which corresponds to $\Delta n = 3.2 \times 10^{-11}$ cm$^{-2}$.

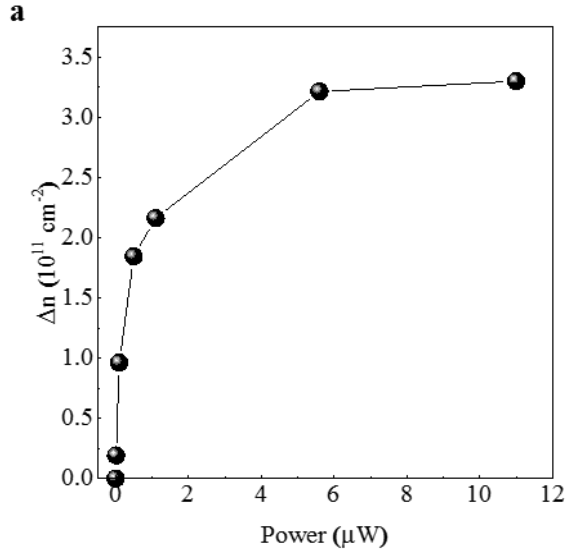

**Figure S5.** (a) Effective photodoping of TSS ($\Delta n$) extracted from Fig. 2a, with the consideration of the effective gate capacitance.

## 6. Weak anti-localization analysis

The analysis of weak anti-localization (WAL) effect (Fig. S6) indicates that only the TSS contribute to the conduction channels in our TI based devices. Because the WAL effect constitutes a prominent transport property of TSS, the parameter α, which extracted from Hikami-Larkin-Nagaoka formula, is essential to differentiating the transport nature of TI and reflects the number of conducting channels[1,2]. As shown in Fig. S6f, α ~ 1 reflects that only two independent conduction channels (top and bottom surface states) exist, insulating bulk state has no contribution to transport under different optical illumination conditions.

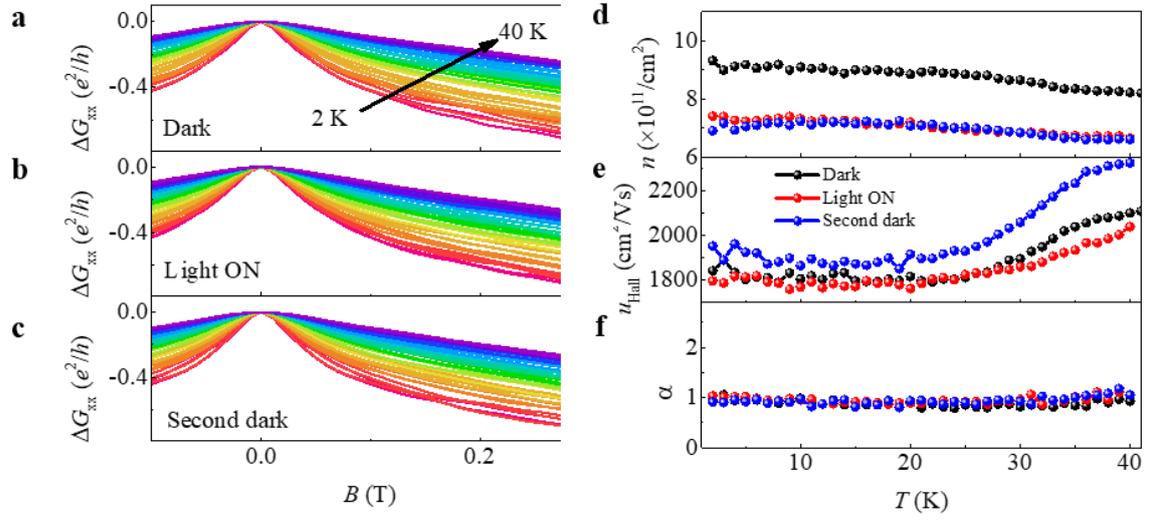

**Figure S6.** (a)-(c) The magnetoconductivity at different temperatures and different illumination conditions. (d) Carrier density extracted from (a)-(c). (d) and (e) Hall mobility extracted from (a)-(c). (f) Coefficient α, extracted from Hikami-Larkin-Nagaoka formula, as a function of temperature.

## 7. Boron nitride gated devices

In order to confirm that the photodoping effect is from bulk state from the TI rather than other extrinsic interfaces, we measured devices with h-BN flake working as the gate dielectric, as h-BN has been shown to be a perfect dielectric material for 2D materials (schematically shown in Fig. S7a and S7d). For h-BN bottom-gated device, as shown in Fig. S7b, the device also shows a large hysteresis in dark (red curve) at 2 K, and the hysteresis disappears with the optical illumination power of 11 μW (purple curve). The data for different initial gate voltages from 60 V to -20 V are shown in Fig. S7c. For h-BN top-gated device, as shown in Fig. S7e and S7f, the hysteresis also disappears under an 11 μW light illumination. All the BN gated TI devices show similar photodoping behavior as what is reported in the main text. It indicates that the photodoping effect is intrinsic for Sn-BSTS.

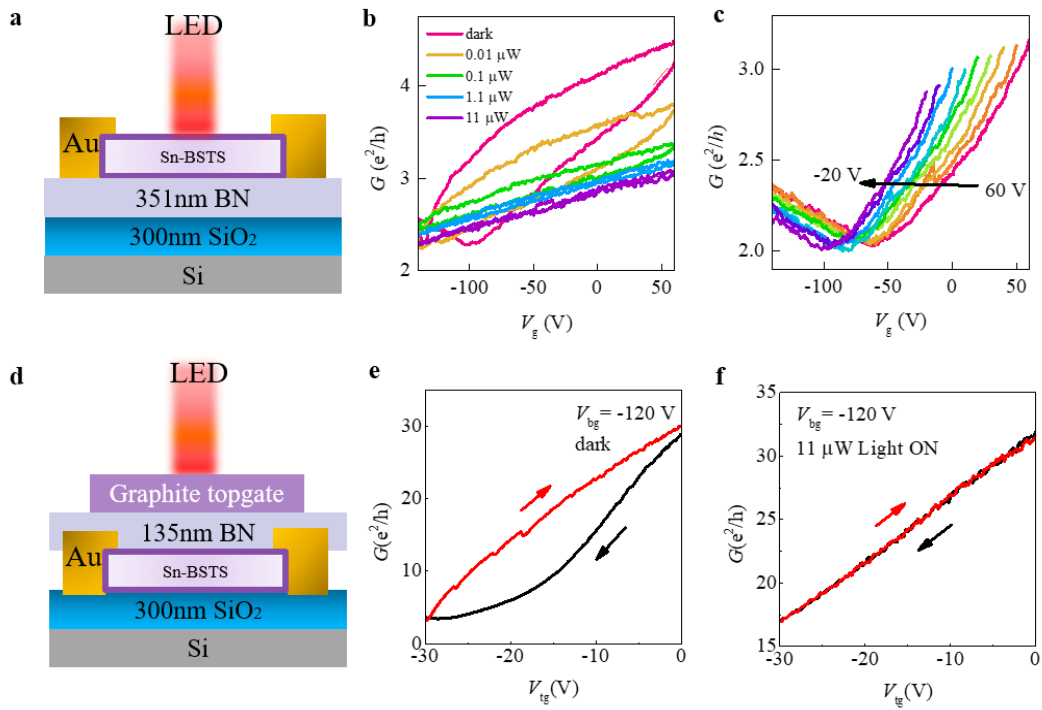

**Figure S7.** (a)-(c) Light-controlled hysteresis loops and nonvolatile memory curves of a TI FET on BN/SiO2/Si substrate. (d)-(f) Light-controlled hysteresis loops of a TI FET on with BN top gate. Devices with different gat voltages show similar hysteresis and photo-doping behaviour.

## 8. Calculation of the Fermi energy of the TSS

Considering the dispersion of the 2D massless Dirac fermions, the density of states of the TSS is given by:

$$D(E) = \frac{|E|}{2\pi\hbar^2 V_F^2},$$

where $E$ is the energy related to the Dirac point, $\hbar$ is the reduced Planck constant, $V_F$ is the Fermi velocity.

Treating the Fermi distribution at 2 K as a step function, the Fermi energy can be related to the carrier density and the gate voltage by:

$$\int_0^{E_F} \frac{E}{2\pi\hbar^2 V_F^2} dE = \frac{E_F^2}{4\pi\hbar^2 V_F^2} = n = c_{eff}|V_g - V_{cnp}|/e,$$

where $E_F$ is the Fermi energy related to the Dirac point, n is the 2D carrier density, $c_{eff}$ is the effective capacitance per unit area coupled to the TSS, $V_g$ is the gate voltage, $V_{cnp}$ is the value of $V_g$ at the charge neutral point (CNP), and $e$ is the electron charge. The Fermi energy at specific $V_g$ is given by:

$$E_F = sign(V_g)\sqrt{4\pi\hbar^2 V_F^2 c_{eff}|V_g - V_{cnp}|/e}$$

To calculate $E_F$, we use $V_F = 2 \times 10^7$ cm/s [3], $V_{cnp} = -5.5$ V.

(1) If the effective capacitance under light illumination is used, meaning $c_{eff} = 0.085\ c_{ox}$, where $c_{ox}$ is the capacitance per unit area of the 300 nm $SiO_2$, we obtain $E_F = 29.5$ meV for $V_g = 60$ V and $E_F = -31.5$ meV for $V_g = -80$V. In this scenario, most of defect sites are empty, and the corresponding $E_F$ corresponds to a lower bound of the highest defect energy.

(2) If the effective capacitance in dark is used, then $c_{eff} = 0.20\ c_{ox}$. We obtain $E_F = 45.3$ meV for $V_g = 60$ V and $E_F = -48.3$ meV for $V_g = -80$V. In this scenario, most of defect sites are occupied, and the corresponding $E_F$ corresponds to a higher bound of the highest defect energy.

Considering these are the values of $V_g$ where the persistent photodoping effect is erased, we estimate the energy distribution of the donor/acceptor levels to be +/- (30 ~ 50) meV away from the Dirac point.